\documentclass{ws-ijmpc}
\usepackage{rotating}

\newcommand{\be}{\begin{equation}}
\newcommand{\ee}{\end{equation}}

\begin{document}

\markboth{J. Mi\'skiewicz, M. Ausloos}
{An attempt to observe economy globalization...} 

\catchline{}{}{}{}{}

\title{An attempt to observe economy globalization: the cross correlation
distance evolution of the top 19 GDP's} 

\author{J. Mi\'skiewicz}

\address{Wroc\l{}aw University, Institute of Theoretical Physics \\ pl. M. Borna
9, 50-204 Wroc\l{}aw, Poland \\ jamis@ift.uni.wroc.pl}

\author{M. Ausloos}

\address{SUPRATECS, B5, University of Li$\grave e$ge, \\ B-4000 Li$\grave e$ge,
Euroland\\ marcel.ausloos@ulg.ac.be}

\maketitle

\begin{history}
\received{Day Month Year}
\revised{Day Month Year}
\end{history}

\begin{abstract}
Economy correlations between the 19 richest countries are investigated through their
Gross Domestic Product increments. A  distance is defined  between increment
correlation  matrix elements and their evolution studied as a function of time and time window size.
Unidirectional and Bidirectional Minimal Length Paths are generated and analyzed for different time windows. A
sort of critical correlation time window is found indicating a transition for best
observations. The mean length path decreases with time, indicating stronger
correlations. A new method for estimating a realistic minimal time window to
observe correlations  and deduce macroeconomy conclusions
from such  features is thus suggested. 

\keywords{corrrelations, gross domestic product, globalization, econophysics, linear network}
\end{abstract}

\section{Introduction}
\label{intro}

How the national economy changes and notably how it increases, i.e. macroeconomic
considerations have entered econophysics research. 

One question which has been raised is whether there is a possibility to model
macroeconomic questions and features from statistical models or better in our
opinion from so-called microscopic models \cite{acp1}${}^-$\cite{acp5}. There have been attempts a long time
ago and more recently about so called business cycles e.g. \cite{refsonbuscycles}.
 There is no certain way with so few data points to claim
that the distribution of GDP increments, or cumulative increments, called
recession or prosperity periods, follow a power law or an exponential
distribution. This has been discussed elsewhere \cite{power_law,power_law1,r2}. A
few papers \cite{r2,r1} of interest pertain to scaling laws in the durations of
recession and prosperity, e.g. when measured through the GDP of a country. Indeed
in most countries, the Gross Domestic Product (GDP) is the official measure of
the economic output because it is the best basis for evaluating the well being of
the citizens of a given country \cite{rbest}. 

GDP is usually defined as the total market value of all final goods and services
produced in a country in a given year, equal to total consumption, investment and
government spending, plus the value of exports, minus the value of imports \cite{gdp_def}.
GDP differs from the Gross National Product (GNP): For example,
in the USA GNP includes the value of goods produced by USA citizens abroad, in
China, Argentina or  Brazil.

There are formidable problems involved in constructing GDP data series, many of
which are discussed in \cite{omer8}. It is important for readers with a natural
science background to appreciate that, outside financial markets, almost all
economic data is estimated and the margin of error around any individual data
point can be substantial. This is the case even with modern estimates of GDP data
for the most recent years. The business activity might, as it is known, be
measured through qualitative factors, including the concept of economic sentiment
indicator, and others sometimes more quantitative; see the interesting studies by \cite{Matkowski}. 
Moreover the time scales are not very clear, e.g it is known
that it takes a long time before some policy change is implemented, has some
effect which can be later analyzed and corrected if necessary. Parameters are
known to be hardly precise, and the variations of variables are not
quantitatively well known. Such macroeconomic studies are thus a challenge to
econophysicists because the set of data points is usually small, and not much is
really proven, - according to usual physicist insight. 

Among (other) classical points raised in macroeconomy and media is the so called
globalization of the economy. In the review \cite{orourke_rev} by O'Rourke and
Williamson \cite{Orourke} one finds some coverage of GDP statistics per capita
and GDP per worker hour, but the authors consider these macroeconomic aggregates
inadequate because they are generally unreliable and the more so frequently not
even available. The main criticisms are perhaps that the most important defect
resides in the averaging of all incomes, whence bypassing much valuable data
needed to understand the factors affecting migration patterns, political
responses to globalization, and the sources of convergence of economics. 

Even though this GDP analysis is considered to be full of defects, it is
sometimes used in order to prove the defects of globalization: see AFL-CIO
statement \cite{aflcio} : {\it Globalization ... has produced a race to the bottom in which companies search the
globe for the lowest possible labor costs and weakest environmental safeguards.
Today's global economy has greatly increased the income gap worldwide, making the
rich even wealthier and eroding working families' standard of living.} 

Others, like \cite{Frankel}, would claim that {\it globalization ... is less
impressive than most non-economists think, judged either by the standard of 100
years ago or by the hypothetical standard of perfect international integration}.
Let us point out that he also considers implications
for economic growth {\it not} measured by GDP. Therefore, it seems unclear
whether economy globalization is proven: it is neither described along
statistical lines nor accounted for by microscopic models.

Thus in order to contribute to these questions, we have performed correlations
studies of GDP, or more precisely annual GDP increments, for the latest 54 years
(1950-2003) as if every country economy can be described by the time series
$y(t)$ of its annual GDP.  
The reason why increments are considered should be stressed at first. While
summation series, like $ \Sigma y(t) $ illustrate the long term development, the
annual increment series emphasizes the fluctuations, i.e. volatility and
variability. In some sense the former is a so called mean field like
approximation approach, sort of moving average, degrading the role of
fluctuations, while the latter emphasizes the possible critical aspects, in a
thermodynamics sense; whence our choice of the signal to be analysed. 

We are not at this time interested in regional or interregional disparities, but
rather care about the leading actors, i.e. the  19 richest countries. They
are in (English) alphabetical order: Austria (AUT), Belgium (BEL), Canada (CAN),
Denmark (DNK), Finland (FIN), France (FRA), Germany\footnote{Germany is
considered as a single country. To have a whole record the data are
constructed as the sum of GDP of both German countries before consolidation.} (DEU), Greece (GRC),
Ireland (IRL), Italy (ITA), Japan (JPN), the Netherlands (NLD), Norway (NOR),
Portugal (PRT), Spain (SPE), Sweden (SWE), Switzerland (CHE), United Kingdom
(GBR) and USA. Conventional (three letter) notations are used for labelling the
countries following the Organization for Economic Co-operation and Development
\cite{oecd}. The GDP yearly increments are next considered and their statistics
are reported in Sect. \ref{dat_anal}.

It is of interest to see whether they are connected  and if in some way how.
We will discuss whether they form clusters, and if these are stable or not. 
 Recall that there are different ways to
find clusters in communities, i.e. through Random Walk (\cite{rnd_walk}),
Laplacian (\cite{laplacian}), Adjacent... matrix eigenvalue and eigenvector
displays. Here we consider a more simple quite pedestrian approach based on
1-dimensional geometric display, emphasizing the strongest connections only.
 In order to do so, a ''distance matrix'' is constructed. The unidirectional (with a given initial
point) and bidirectional minimal length paths (UMLP and BMLP respectively) are
next constructed, as a function of time and for moving time windows, having various
(but constant during the displacement) sizes.  

The "local correlations" between these GDP {\it increments} are then investigated
as follows. The evolution of the mean distance between countries are 
reported as a function of time for 5y, 15y, 25y and 35y time windows. 
It is found that different time windows emphasize different features, in particular the "strength of correlations". A
sort of critical correlation time (window) is found indicating a transition for
observations {\it and} correlation strengths. A new method for estimating a
realistic minimal time window to observe meaningful correlations in such
macroeconomy features and questions is thus suggested. In so doing, clustering
effects from the point of view of correlations between the GDP  of the 19 most developed countries and globalization of these economies are observed. 

It is thereby obtained that an economic conclusion can be derived through such a
statistical physics approach: an increase in correlations, seen through a
reduction of the UMLP and BMLP overall length for the GDP annual increments of
such rich countries, is now a proven feature. 

\section{Data source}
\label{dat_source}

The GDP of the 19 richest countries are taken from the web \cite{websource}.
The GDP values for each of these countries
are first normalised to their 1990 value given in US dollars as published by the
Groningen Growth and Development Center on their web page \cite{ggdc}. The data
covers the period between 1950 and 2003, i.e. 54 points for each country. An
additional "All" entity has been added for comparison. Its GDP value is
constructed as the sum of 19 country GDP. This All "artificial country" can serve
as some base (more precisely "top") line for globalization idea reference. In
fact it is also possible to introduce a normalised increment as a increment with
respect to that of ''All'', but for the sake of simplicity we do not exploit this
idea.

\section{Data analysis}
\label{dat_anal}

The yearly GDP increments to be studied are definned as
\be
\Delta GDP(t) = \frac{GDP(t) - GDP(t-1)}{GDP(t-1)} 
\ee
calculated at the end of a year $t \in (1951; 2003 ) $.

We have tested the normality of the GDP increment through the 
Jarque-Bera ($JB$) test \cite{jb_stat}. First the $JB$ coefficient is defined as 
\be
\label{jb_test}
JB=\frac{T}{6} (S^2 +\frac{(K-3)^2}{4}),
\ee
where $T$ denotes  the number of data points. Notice that the Jarque-Bera test for normality takes
into account two distribution parameters: skewness ($S$) and kurtosis ($K$). Then the value of $JB$ is
compared with the $\chi^2$ statistics with 2 degrees of freedom. The value of 5\%
confidence level from a $\chi^2$ statistics with 2 degrees of freedom is 5.99.
It is found that only Switzerland does not satisfy  such a normality test, i.e.
$JB_{CHE} =6.57$. In the case of Germany, All and Ireland the GDP increment
statistics is highly Gaussian ($JB_{DEU} =0.269$, $JB_{All} =0.501$, $JB_{IRL}
=0.715$). The second set of countries with highly Gaussian statistics of GDP
increments ($JR \in (1,1.2)$) are: USA, FRA, ITA, NLD, BEL, SWE, GRC. 

\subsection{Distance matrices}

The distance between countries is defined following \cite{mantegna01} 
\be
\label{odl_01}
d(i,j)_{(t,T)} = \sqrt{2(1- \Gamma_{(t,T)} (c_i,c_j))} 
\ee
where the correlation function is defined as 
\be
\Gamma_{(t,T)} (c_i,c_j) = \frac{<c_i
c_j>_{(t,T)} - <c_i>_{(t,T)} <c_j>_{(t,T)} }{\sqrt{(<c_i^2>_{(t,T)} -
<c_i>^2_{(t,T)}) (<c_j^2>_{(t,T)} - <c_j>^2_{(t,T)}})}, 
\ee
where $ c_i $ denotes the time series of increments of GDP for the $i^{th}$
country, and $ <c_i>_{(t,T)} $ is the average of yearly GDP increments in the
time window $ (t,t+T)$ of size $ T$.

An example  of such a correlation matrix is reported for
illustration purpose in the case of the shortest (5y)  time window ending in
2003 in Table \ref{odl_tab05}.

{\small
\begin{sidewaystable}
\begin{center}
\tbl{The distances between countries in a time window of 5y}
{\begin{tabular}{@{}lcccccccccccccccccccc@{}} 
\toprule & \tiny AUT & \tiny BEL & \tiny CAN & \tiny DNK & \tiny FIN & \tiny FRA
& \tiny GRC & \tiny IRL & \tiny ITA & \tiny JPN & \tiny NLD & \tiny NOR & \tiny
PRT & \tiny SPE & \tiny SWE & \tiny CHE & \tiny GBR & \tiny USA & \tiny DEU &
\tiny All \\ \hline \tiny AUT & \tiny 0 & \tiny 0.694 & \tiny 0.726 & \tiny 0.557
& \tiny 0.235 & \tiny 0.483 & \tiny 1.63 & \tiny 0.754 & \tiny 0.92 & \tiny 1.4 &
\tiny 0.42 & \tiny 0.617 & \tiny 0.418 & \tiny 0.412 & \tiny 0.534 & \tiny 0.522
& \tiny 0.474 & \tiny 0.502 & \tiny 0.548 & \tiny 0.438 \\ \hline \tiny BEL &
\tiny 0.694 & \tiny 0 & \tiny 0.421 & \tiny 0.470 & \tiny 0.686 & \tiny 0.486 &
\tiny 1.4 & \tiny 0.386 & \tiny 0.701 & \tiny 0.998 & \tiny 0.658 & \tiny 0.709 &
\tiny 0.704 & \tiny 0.552 & \tiny 0.375 & \tiny 0.581 & \tiny 0.459 & \tiny 0.684
& \tiny 0.338 & \tiny 0.396 \\ \hline \tiny CAN & \tiny 0.726 & \tiny 0.421 &
\tiny 0 & \tiny 0.317 & \tiny 0.724 & \tiny 0.755 & \tiny 1.57 & \tiny 0.221 &
\tiny 1.08 & \tiny 1.20 & \tiny 0.725 & \tiny 1.00 & \tiny 0.820 & \tiny 0.688 &
\tiny 0.226 & \tiny 0.880 & \tiny 0.739 & \tiny 0.418 & \tiny 0.68 & \tiny 0.364
\\ \hline \tiny DNK & \tiny 0.557 & \tiny 0.470 & \tiny
0.317 & \tiny 0 & \tiny	0.471 & \tiny 0.717 & \tiny 1.5 & \tiny 0.491 & \tiny
1.02 & \tiny 1.16 & \tiny 0.737 & \tiny	0.932 & \tiny 0.802 & \tiny 0.688 &
\tiny
0.313 & \tiny 0.739 & \tiny 0.580 & \tiny 0.409 & \tiny 0.632 & \tiny 0.16 \\
\hline \tiny FIN & \tiny 0.235 & \tiny 0.686 & \tiny 0.724 & \tiny 0.471 &
\tiny	0
& \tiny 0.587 & \tiny 1.51 & \tiny 0.807 & \tiny 0.913 & \tiny 1.28 & \tiny 0.625
& \tiny 0.701	& \tiny 0.623 & \tiny 0.587 & \tiny 0.578 & \tiny 0.509 & \tiny
0.421 & \tiny 0.568 & \tiny 0.586 & \tiny 0.383 \\ \hline \tiny FRA & \tiny 0.483
& \tiny 0.486 & \tiny 0.755 & \tiny 0.717 & \tiny 0.587 & \tiny 0 & \tiny 1.52 &
\tiny 
0.673 & \tiny 0.556 & \tiny 1.23 & \tiny 0.362 & \tiny 0.281 & \tiny	0.321
& \tiny 0.234 & \tiny 0.571 & \tiny 0.32 & \tiny 0.346 & \tiny 0.756 & \tiny
0.186 & \tiny 0.568 \\ \hline \tiny GRC & \tiny 1.63 & \tiny 1.4 & \tiny 1.57 & \tiny
1.5 & \tiny 1.51 & \tiny 1.52 & \tiny 0 & \tiny 1.62 & \tiny 1.14 & \tiny
0.521 & \tiny 1.73 & \tiny 1.44 & \tiny 1.7 & \tiny 1.65 & \tiny 1.61 & \tiny 1.34
& \tiny 1.35 & \tiny 1.74 & \tiny 1.42 & \tiny 1.51 \\ \hline \tiny IRL & \tiny
0.754 & \tiny 0.386 & \tiny 0.222 & \tiny 0.911 & \tiny 0.807 & \tiny 0.673 &
\tiny 1.61 & \tiny 0 & \tiny 1.02 & \tiny 1.24 & \tiny 0.628 & \tiny	0.928 &
\tiny 0.729 & \tiny 0.592 & \tiny 0.248 & \tiny	0.866 & \tiny 0.752 & \tiny 0.491
& \tiny 0.616 & \tiny 0.472 \\ \hline \tiny ITA & \tiny 0.92 & \tiny 0.701 &
\tiny 1.08 & \tiny 1.02 & \tiny	0.913 & \tiny 0.556 & \tiny 1.14 & \tiny 1.02
& \tiny 0 & \tiny 0.885 & \tiny 0.885 & \tiny	0.435 & \tiny 0.823 & \tiny 0.773 &
\tiny 0.978 & \tiny 0.437 & \tiny 0.533 & \tiny 1.198 & \tiny 0.496 & \tiny 0.915 \\
\hline \tiny JPN & \tiny 1.4 & \tiny 0.998 & \tiny 1.20 & \tiny 1.16 & \tiny 1.28
& \tiny 1.23 & \tiny 0.521 & \tiny 1.24 & \tiny 0.885 & \tiny 0 & \tiny 1.47 &
\tiny 1.23 & \tiny 1.47	& \tiny 1.38 & \tiny 1.27 & \tiny 1.09 & \tiny
1.05	& \tiny 1.46 & \tiny 1.09 & \tiny 1.18 \\ \hline \tiny NLD & \tiny 0.42 & \tiny 0.658 &
\tiny 0.725 & \tiny 0.737 & \tiny 0.625 & \tiny 0.362 & \tiny 1.72 & \tiny 0.628 
& \tiny 0.885 & \tiny 1.47 & \tiny 0 & \tiny	0.544 & \tiny 0.127 & \tiny 0.136
& \tiny 0.518 & \tiny 0.625 & \tiny 0.621 & \tiny 0.572 & \tiny 0.502 & \tiny
0.61 \\ \hline \tiny NOR & \tiny 0.617 & \tiny 0.709 & \tiny 1.00 & \tiny 0.932 &
\tiny 0.70 & \tiny 0.281 & \tiny 1.44 & \tiny 0.928 & \tiny 0.435 & \tiny 1.23 & \tiny
0.544 & \tiny 0	& \tiny 0.447 & \tiny 0.457 & \tiny 0.830 & \tiny 0.297 &
\tiny 0.441 & \tiny 0.980 & \tiny 0.385 & \tiny 0.790 \\ \hline \tiny PRT & \tiny 0.418
& \tiny 0.704 & \tiny 0.820 & \tiny 0.802 & \tiny 0.623 & \tiny 0.321 & \tiny 1.7
& \tiny 0.729 & \tiny 0.823 & \tiny 1.46 & \tiny 0.127 & \tiny	0.447 & \tiny 0
& \tiny 0.163 & \tiny 0.613 & \tiny 0.563 & \tiny 0.592 & \tiny 0.67 & \tiny
0.489 & \tiny 0.664 \\ \hline \tiny SPE & \tiny 0.412 & \tiny 0.552 & \tiny 0.688 &
\tiny 0.688 & \tiny 0.587 & \tiny 0.234 & \tiny 1.65 & \tiny 0.592 & \tiny 0.773
& \tiny 1.38 & \tiny 0.136 & \tiny	0.457 & \tiny 0.163 & \tiny 0 & \tiny 0.480
& \tiny 0.516 & \tiny 0.507 & \tiny 0.603 & \tiny 0.369 & \tiny 0.546 \\ \hline \tiny SWE
& \tiny 0.534 & \tiny 0.375 & \tiny 0.226 & \tiny 0.313 & \tiny 0.578 & \tiny
0.571 & \tiny 1.61 & \tiny 0.248 & \tiny 0.978 & \tiny 1.27 & \tiny 0.518 & \tiny
0.830 & \tiny 0.613 & \tiny 0.480 & \tiny 0 & \tiny 0.728 & \tiny 0.600 & \tiny
0.325 & \tiny 0.526 & \tiny 0.255 \\ \hline \tiny CHE & \tiny 0.522 & \tiny 0.581
& \tiny 0.880 & \tiny 0.739 & \tiny 0.509 & \tiny 0.32 & \tiny 1.34 & \tiny 0.866 & \tiny
0.437 & \tiny 1.085 & \tiny 0.625 & \tiny	0.297 & \tiny 0.563 & \tiny 0.516
& \tiny 0.728 & \tiny 0 & \tiny 0.175 & \tiny 0.892 & \tiny 0.296 & \tiny 0.608
\\ \hline \tiny GBR & \tiny 0.474 & \tiny 0.459 & \tiny 0.739 & \tiny 0.580 &
\tiny 0.421 & \tiny 0.346 & \tiny 1.35 & \tiny 0.752 & \tiny 0.533 & \tiny 1.05 &
\tiny 0.621 & \tiny	0.441 & \tiny 0.592 & \tiny 0.507 & \tiny 0.600 & \tiny 0.175
& \tiny 0 & \tiny 0.782 & \tiny 0.255 & \tiny 0.453 \\ \hline \tiny USA & \tiny
0.502 & \tiny 0.684 & \tiny 0.418 & \tiny 0.409 & \tiny 0.568 & \tiny 0.756 & \tiny
1.74 & \tiny 0.491 & \tiny 1.2 & \tiny 1.5 & \tiny 0.572 & \tiny 0.980 & \tiny
0.67 & \tiny 0.603 & \tiny 0.325 & \tiny 0.892 & \tiny 0.782 & \tiny 0 & \tiny
0.766 & \tiny 0.407 \\ \hline \tiny DEU & \tiny 0.548 & \tiny 0.338 & \tiny 0.676
& \tiny 0.632 & \tiny 0.586 & \tiny 0.186 & \tiny 1.42 & \tiny 0.616 & \tiny
0.496 & \tiny 1.09 & \tiny 0.502 & \tiny	0.385 & \tiny 0.489 & \tiny 0.369 &
\tiny 0.526 & \tiny 0.296 & \tiny 0.255 & \tiny 0.766 & \tiny 0 & \tiny 0.494 \\ \hline
\tiny All & \tiny 0.438 & \tiny 0.396 & \tiny 0.364 & \tiny 0.16 & \tiny	0.383 & \tiny
0.568 & \tiny 1.51 & \tiny 0.472 & \tiny 0.915 & \tiny 1.18 & \tiny 0.607 & \tiny
0.790 & \tiny 0.664 & \tiny 0.546 & \tiny 0.255 & \tiny 0.608 & \tiny 0.453 &
\tiny 0.407 & \tiny 0.494 & \tiny 0 \\ \botrule 
\end{tabular}
\label{odl_tab05} }
\end{center}
\end{sidewaystable}

Other possibilities to define a distance between time series are found in the
literature \cite{mantegna01,mantegna02,lee} but are not
considered here, because such definitions sometimes loose the information about
the difference between correlated and anticorrelated time series. 

\subsection{MLP algorithms}

In order to obtain some quantitative information on the country correlations we
have looked for clustering possibilities. There are several classical ways to
display clusters through graph ordering or network construction techniques, like
through the so called Minimum Spanning Tree, which has indeed found many
illustrations in stock markets analysis \cite{mst} and GDP correlation 
studies\cite{ortega}. However the MST is far
from univocal. Moreover the present MST would be of very limited size. If the
MST has a homogeneous node distribution, its ''diameter'' should be of the order
of $ln$ $N$, i.e. $\simeq 3 $, while it should be of order $\sqrt N  \simeq 4.47$
if it is a causal tree \cite{havlin}.
In any case, the present trees will have very few branches, and at most four
levels. Not much structure would show up. Whence we discarded constructing a
MST in our investigations and opted for an apparently more pedestrian one
dimensional (1-D) approach.
The following minimal length path algorithms (MLP)  emphasize the { \it strongest} correlation
between entities through the constraint that the item is attached only once to the
network. This results in a lack of loops in the ''tree''. The construction of
more elaborate networks is left for further studies. 

Two different minimal length path (MLP) algorithms have been developed 
and analyzed, i.e.  the so called unidirectional (UMLP) and bidirectional (BMLP)
paths.
\begin{description}
\item[UMLP] The algorithm begins with choosing an initial point as a seed. Here
the initial point is the "All" country. Next the shortest connection (in terms of
the distance definition, Eq.(\ref{odl_01}) is looked for between the seed and the
other possible neighbors. The closest possible one is selected and attached to
the seed. One searches next in the matrix for the
entity closest to the previously attached one, and repeat the process.
\item[BMLP] The algorithm begins with searching for the pair of countries which
has the shortest distance between them. Then these countries become the root of a
chain. In the next step the closest country for either ends of the chain is
searched. The shortest distance being found, the country is attached to the
appropriate end. Next a search is made for the closest neighbor of the new ends
of the chain. Being selected, the entity is attached, a.s.o. Notice that there is
some arbitrariness in the choice of the country position in the initial bond. We
have chosen the "alphabetical" order from left to right or bottom to up, for all
displayed graphs, i.e. the same order as on the example of distance matrix
(Table \ref{odl_tab05}). 
\end{description}

In view of the UMLP and BMLP definitions, it is obvious that UMLP expands only in
one direction while BMLP may also grow in opposite ones. 

The UMLP and BMLP chains have been constructed for all possible time windows. For
illustration purpose both cases are displayed, each year, for the 5y time window
(Figs \ref{odl_05}). Recall that the last $t$ value corresponds to Dec. 31, 2003.
Notice that the first data point on the time axis in each figure depends on the
time window considered. 
The UMLP and BMLP lengths vary with time for a given time window. E.g. for the 5y
time window the UMLP and BMLP extends up to 16 or so, but has a minimum near 7.5.

The total length of a chain can be interpreted as a measure of the strength of
correlations between countries: the more compact is a chain, the greater are the
correlations. Therefore the chain properties can be used as a quantitative
measure of a globalization process. It can be also searched whether the
correlations are changing significantly with time.

\begin{figure}[ht]
\centerline{\psfig{file=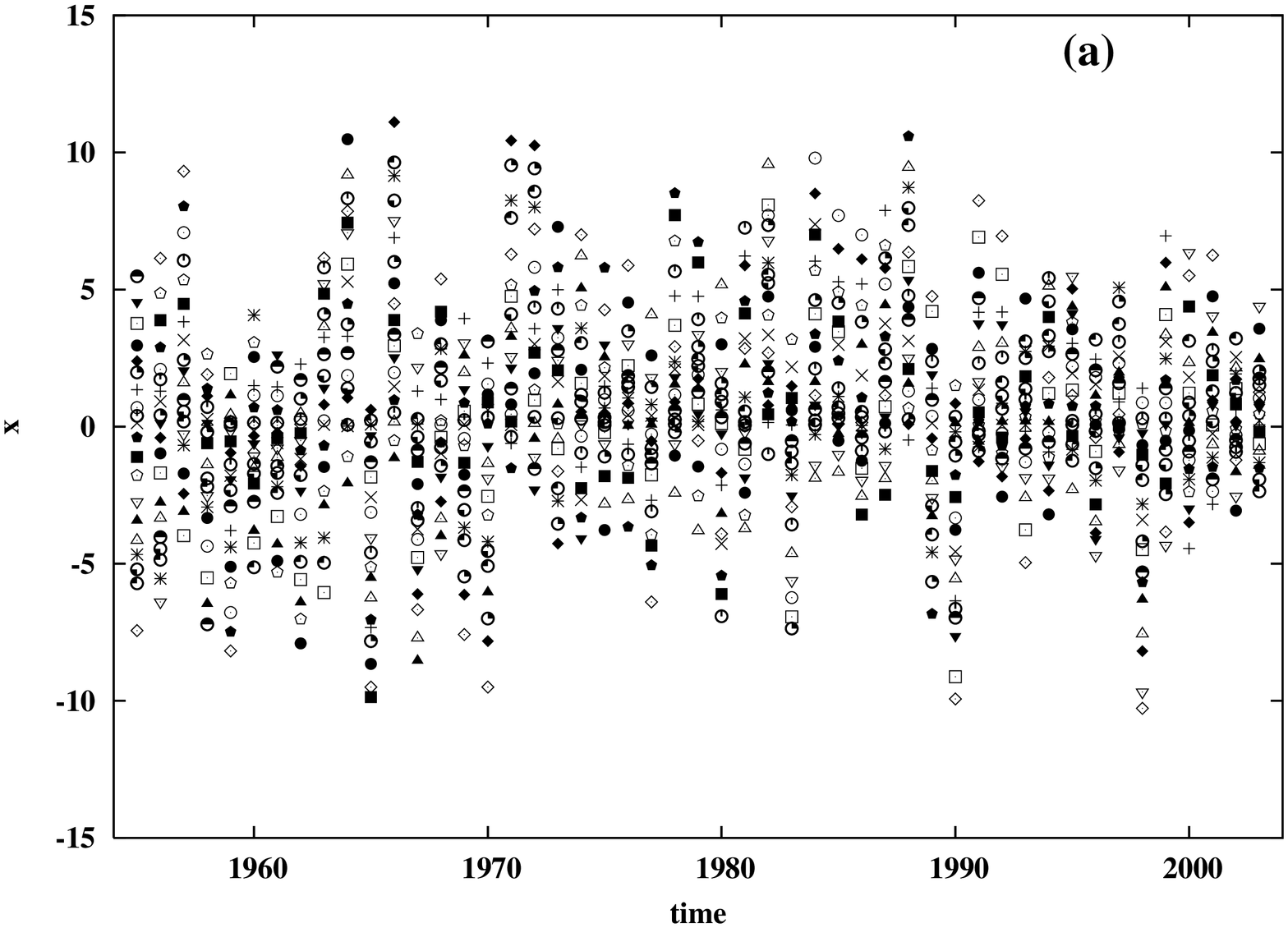, scale=0.5, angle =-90}}
\centerline{\psfig{file=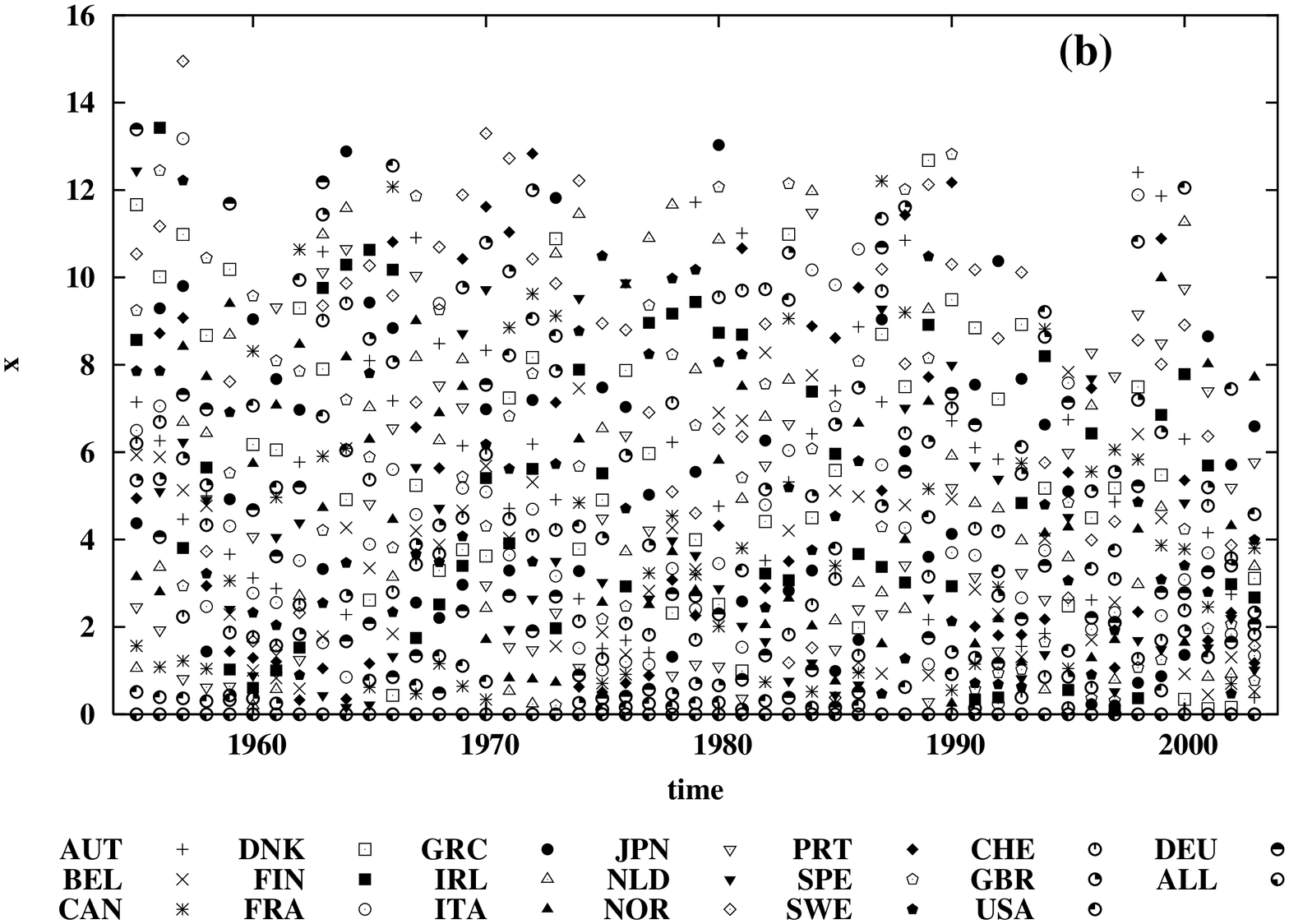, scale=0.5, angle =-90}}
\caption{\label{odl_05} Position of countries as a function of time in the case
of (a) BMLP and (b) UMLP algorithm for a time window of 5y } 
\end{figure}

\begin{figure}
\centerline{\psfig{file=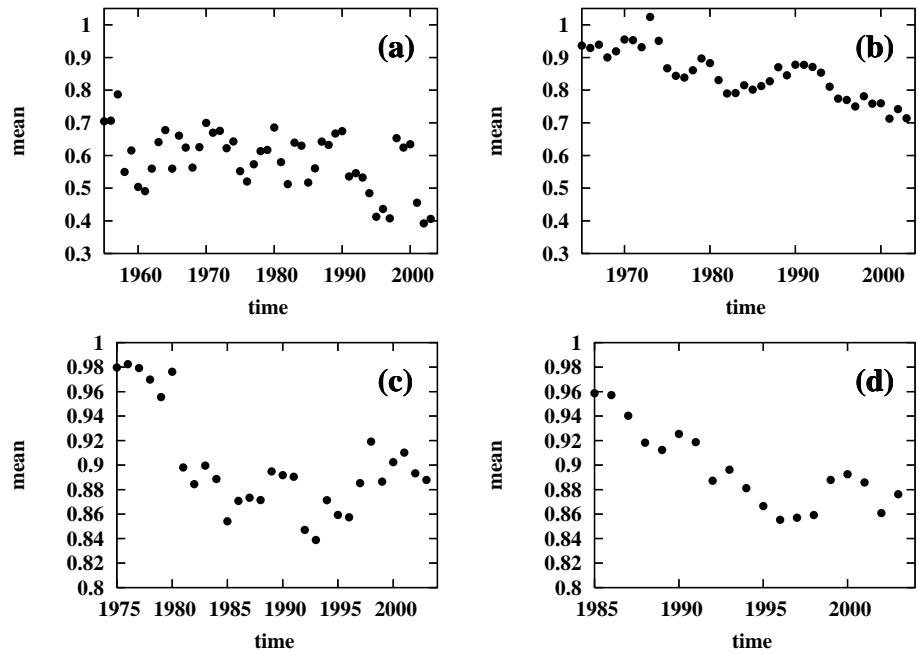, scale=0.5, angle =-90}}
\caption{Mean distance between countries in the case of UMLP algorithm for the moving time
windows: (a) 5y, (b) 15y, (c) 25y, (d) 35y} \label{fig:chain_end_mean_a}
\end{figure}

\begin{figure}
\centerline{\psfig{file=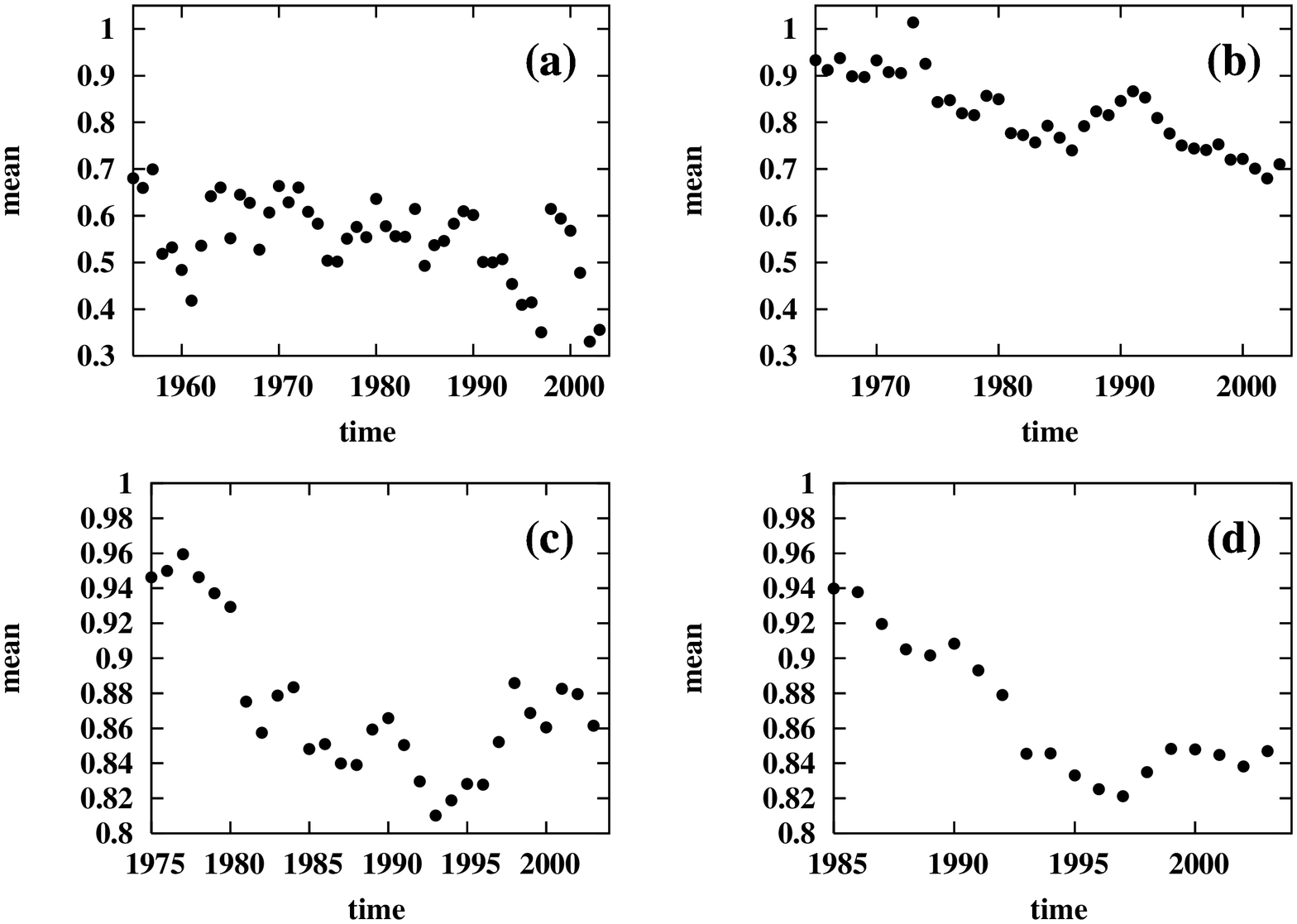, scale=0.5, angle =-90}}
\caption{Mean distance between countries in the case of BMLP algorithm for the moving time
windows: (a) 5 yrs, (b) 15 yrs, (c) 25 yrs, (d) 35 yrs}
\label{fig:chain_end_mean_b}
\end{figure}

\section{Minimal length path analysis: time window observed effects} 

The position of a country in the MLP chain represents the relative distance
between countries with respect to the measure, Eq. \ref{odl_01}. So both UMLP
and BMLP chain algorithms rank the correlations between countries with respect to
the relative distance.

The mean   of each chain has been calculated in every possible year for every
possible time window (Figs. \ref{fig:chain_end_mean_a}- \ref{fig:chain_end_mean_b}).

\subsection{5 years time window}
\label{5_lat}

Considering the position of a country in a BMLP or UMLP chains 
(Figs. \ref{odl_05} (a-b)) it can be found
that within the 5y time window the position of countries is rapidly changing for
both BMLP and UMLP algorithms.
Therefore it is difficult to distinguish any specific or regular behavior 
in such a time window. Generally the distances are decreasing (from 1.0 to 0.6)
obtaining the lowest values at the end of the considered time interval (1995 -
2003) with the exception for the interval 1998 - 2000.

\subsection{15 years time window}
\label{15_lat}

In the case of the short time scale, i.e. 15 y, the 
 mean distance and the total length of
the chains do not change significantly as the time window is moved along the time
axis. In comparison to the 5y time window it can be said that the distribution of
distances is stabilising. From the statistical mean value point of view, three
regions can be distinguished for both path construction procedures. In the case
of UMLP, Fig. \ref{fig:chain_end_mean_a}.b, for 1965 - 1982, the mean distance is
decreasing, starting from the value $\approx 1$ and achieving $\approx 0.8$ at
its local minimum. Later on the mean distance is increasing till 1990 before a
decreasing tendency till 2003. The BMLP chain follows the same pattern (Fig
\ref{fig:chain_end_mean_b}.b).

It is worth noticing that within this time window a globalization process can be
already observed. The chains shorten, because the correlations between countries
become stronger, especially in the recent years. 

This conclusion is also supported by a detailed analysis of each country position
in the UMLP and BMLP chains. In the case of the UMLP chain the country closely
related to the average is USA, which is the second country in the chain (just
after ''All'' in 32 out of 39 considered time windows (1965 - 1979, 1987 - 2001
and 2003). The second country is Germany, which five times takes the second
country in the chain (1982 - 1986).
Along this UMLP procedure the strongest correlations are found between USA - All
and USA - All - DEU, whence displaying the leading role of USA and Germany for
the world economy. Besides this set two more pairs can be pointed out: SPE - PRT
(1993 - 2003), and strangely JPN - GRC (1976 - 1988)again. USA, CAN and GBR
remain at very stable positions throughout the whole considered period. In other
periods DEU (1975 - 1979, 1981 - 1984, 1995 - 2001), SWE (1965 - 1975, 1991 -
2001) and BEL (1990 - 1993, 1994 - 2002) do not change much their position in the
BMLP chain.

In the case of BMLP, similar patterns are revealed and strong correlation between
All - USA and USA - All - DEU. JPN - GRC and SPE - PRT is observed. This analysis
shows also an important role of NLD, which is situated very close to the initial
pair of the chain (1975 - 1990). The countries with stable positions in the chains are
CAN (1965 - 1969, 1976 - 1979, 1987 - 1997), FRA (1965 - 1969, 1975 - 1979 and
1987 - 1997).

\subsection{25 years time window}
\label{25_lat}

 As compared with the previously considered cases Sect. \ref{5_lat} and Sect. \ref{15_lat} it
can be noticed that the correlations between countries are stabilising over 
the medium (25y) time scale.
In the mean distance evolution (Figs. \ref{fig:chain_end_mean_a} and
\ref{fig:chain_end_mean_b}) four regions can be distinguished. The mean distance
between countries is decreasing in years (1975 - 1988 and 1990 - 1993), (1975 -
1985 and 1990 - 1993) in the case of BMLP and UMLP respectively. The minimum
value are 0.86 (1995), 0.86 (1993) and 0.84 (1988) and 0.81 (1993) for BMLP and
UMLP chains.

As it was observed for the 15y time window, the most influent countries are USA,
DEU, CAN. However new countries can be added to this set. In BMLP analysis a
cluster (the set of most highly correlated countries) formation can be noticed.
Specially strong and stable correlations can be seen between USA - All - GBR -
CAN in (1975 - 1982), FRA - NLD - DEU - GBR - CAN (1985 - 1992) and ITA - DEU -
All - USA - CAN - GBR (1995 - 2003). The second set of cluster is formed by
countries, which are not placed at the beginning of the chain, but also reveal
strong and stable correlations: NLD - PRT - SPE, JPN - GRC - ITA and ITA - FRA. 

In the case of UMLP the clustering process is not so well seen, especially in the
initial part of the chain, since the initial point has been arbitrarily chosen.
Therefore only strongest correlations can be seen e.g. JPN - GRC - ITA. However
UMLP shows relative position of a country and it can be seen that in many cases
the position is very stable e.g. USA, CAN - almost all considered interval, DEU
(1986 - 1990), GBR (1976 - 1980, 1994 - 2001), FRA (1975 - 1990), IRL (1975 -
1998), NOR (1975 - 1995).

\subsection{35 years time window}

In the long (35 y) time scale the 
 mean distance between countries, and the total
length of the chain, is decreasing through almost all considered time intervals
(1985 - 1997 and 2000 - 2003): the minimum being 0.82 (1997) and 0.85 (1996) for
BMLP and UMLP respectively. This indicates that for the 25y time window analysis,
the globalisation process is observed.

Several clusters are also seen in this long time window (with some small
modifications): BEL - JPN - FRA - NLD - All - USA - CAN (1985 - 1991), GBR - USA
- All - JPN - FRA - BEL - NLD - DEU (1992 - 1994), SPE - FRA - BEL - NLD - DEU -
All - USA (1995 - 2001), GBR - USA - All - DEU - FRA - BEL (2002 - 2003). Besides
those, strong correlations can be seen between JPN - ITA - GRC - AUT (1998 -
2001), JPN - GRC (1085 -1990, 1995 - 1997). In the case of UMLP not only the
special role of USA can be seen but also CAN, BEL, FRA, NLD and DEU, which are
very close to the initial point of BMLP chains in almost all considered cases.
The surprisingly distinct position of DEU (close to the end of chains) in the
interval (1985 - 1992) can be likely explained by the fact, that Germany (DEU)
until Berlin wall fall was two separated countries, but as it was mentioned in
Sect. \ref{intro} it is treated here as a single previously lasting entity. It is
likely though that the definition of GDP might have been diffeent in East and
West Germany before the Berlin wall overthrow.

\subsection{45 years time window}

In the case of very long (45y) time windows there is no point to perform a
detailed statistical analysis, since there are only a few data points available.
However it is worth noticing that both path analyses point out to an undergoing
globalization process: the mean distance is decreasing throughout all considered
moving time windows. The relationship between countries remains similar as in the
previous time window sizes; the leading countries are USA and CAN. An interesting
case is presented by BEL and NLD, which are rather small countries, but their
positions in the ranking is very close to the initial point of a chain. 

This is likely due to the fact, that the economies of those countries are closely
related to the neighboring and leading countries, i.e. much depend on the
economic evolution elsewhere. 

\section{Time scale analysis}
The statistical properties of the averaged distances between countries for the
different time windows located at a given time have been next analysed. The average
distance and average standard deviation for BMLP and UMLP are presented in Fig.
\ref{mean_tot}. The average distance has a downward trend with the increasing
size of the time window, though 7y oscillations seem to be revealed in the UMLP
case. The standard deviation is decreasing (Fig. \ref{mean_tot}) with the size of
the time window for both algorithms.
Three regions can be distinguished. A special role for the 25y time window has
been observed, at which the mean value reaches a relatively stable value
and for which the standard deviation  has a plateau. 

\begin{figure} 
\centerline{\psfig{file=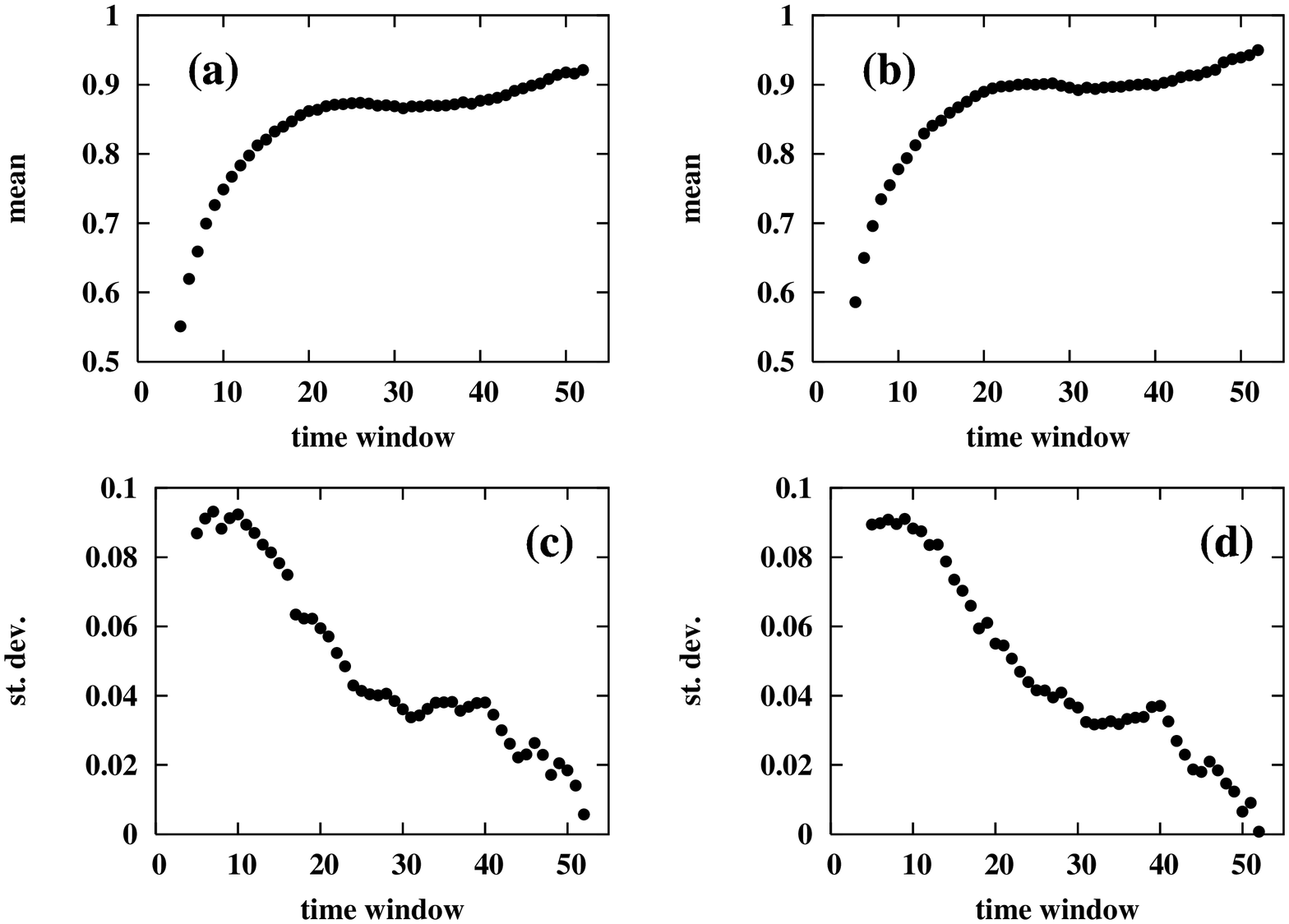, scale=0.55, angle=-90}}
\caption{Mean distances between countries for different time windows, (a) mean value of distances for BMLP algorithm,(b) mean value of distances for UMLP algorithm, (c) standard deviation of distances between countries for BMLP algorithm, (d) standard deviation of distances between countries for UMLP algorithm} 
\label{mean_tot} 
\end{figure}

\section{Conclusion}

The economy evolution of the top 19 richest countries has been investigated
through their Gross Domestic Product increments correlations. The distance
(between increment correlation) matrix has been calculated as a function of time. The Unidirectional and Bidirectional
Minimal Length Paths (UMLP and BMLP)  have been generated and analysed for different time
windows. The total chain length decreases
as a function of time. The choice of a relevant time window is emphasized
for getting less noisy results. 
A sort of critical correlation time window has been found. Comparing the
special role of 25y time window seen from the distances analysis for changing
length of time window with the fact that the correlations between countries are
well seen in the 25y time window (Sect. \ref{25_lat}) it seems that in the case
of investigation of correlation on the world level the 25y time window is the
most appropriate one. Indeed the analysis of UMLP and BMLP graphs shows that the
shortest time window, which allows to observe correlations between countries
should not be shorter than 15y and the most appropriate is of the length of 25y.
Otherwise only the strongest correlations can be observed. Thus such an
observation window also indicates a transition for observations, related to weak
and strong fluctuation correlations, as at a physical phase transition. A new
method for estimating a realistic minimal time window to observe correlations in
such macroeconomy, but also stock market features is thus suggested. 

Because the GDP increment distance between countries is overall decreasing, this
suggests similarities in development patterns, which likely result from
interactions of economies, in a globalization sense. Two at first bizarre cases
can serve as an argument and should be
pointed out. One is the case of Belgium, with a very entangled economy according
to standard political considerations.  
Another is the surprising connection Greece-Japan. Yet, it is known that half the
Japanese fleet is greek owned \cite{Nakata}. 

Notice that when the properties of UMLP and BMLP algorithms are compared, it can
be pointed out, that BMLP is more sensitive to searching for a clustering patterns among the
considered entities, while UMLP is more suitable for ranking countries. In so
doing it could be useful in solving portfolio problem optimizations. 

\section{Acknowledgement} This work was  partially financially supported by FNRS
convention FRFC 2.4590.01. JM would like also to thank SUPRATECS for the
welcome and hospitality. He would like to thank Prof. W. Kwasnicki who has helped
us to find the data source. MA would like to thank Prof. H. Nakata and Prof. E.
Haven for comments and encouragements.

\end{document}